# The first detection of an atmosphere on a trans-Neptunian object beyond Pluto

Ko Arimatsu[1,2], Fumi Yoshida[3,4], Tsutomu Hayamizu[5], Satoshi Takita[6], Katsumasa Hosoi[7], Takafumi Ootsubo[3], and Jun-ichi Watanabe[1]

[1] National Astronomical Observatory of Japan, 2-21-1 Osawa, Mitaka, Tokyo 181-8588, Japan

[2] The Hakubi Center/Astronomical Observatory, Graduate School of Science, Kyoto University, Yoshida-honmachi, Sakyo-ku, Kyoto, 606-8501 Japan

[3] University of Occupational and Environmental Health, Japan, 1-1 Iseigaoka, Yahatanishi, Kitakyushu, Fukuoka 807-8555, Japan

[4] Planetary Exploration Research Center, Chiba Institute of Technology, 2-17-1 Tsudanuma, Narashino, Chiba 275-0016, Japan

[5] Saga Hoshizora Astronomy Center, 328 Nishiyoka-cho, Oaza Takataro, Saga, Saga 840-0036, Japan

[6] Institute of Astronomy, Graduate School of Science, The University of Tokyo, 2-21-1 Osawa, Mitaka, Tokyo 181-0015, Japan

[7] Japan Occultation Information Network (JOIN), Japan

**Trans-Neptunian objects (TNOs) in the outer Solar System are predominantly small, icy worlds long presumed to be atmosphereless except for the largest bodies[1–3]. Until now, Pluto has been unique among TNOs[4,5] in exhibiting a substantial atmosphere (nitrogen with trace methane and carbon monoxide) at microbar pressure levels[6,7]. All other known TNOs, including ~ 1000-km-sized bodies such as Eris, Haumea, Makemake, and Quaoar, have shown no detectable atmospheres in stellar occultation observations[8 – 11], with surface pressure upper limits of order 1–100 nanobars. Here we report the first detection of an atmosphere around a TNO besides Pluto. A stellar occultation by the ~ 250-km-radius plutino (612533) 2002 $XV_{93}$ on 10 January 2024 revealed a refractive signature, indicating the presence of a thin atmosphere. The derived surface pressure is 100–200 nanobars, i.e. approximately a hundred times lower than Pluto's and yet significantly above previous limits for other larger bodies. This discovery provides the first evidence that even a sub-1000-km TNO can retain an atmosphere, challenging current paradigms of volatile retention[1,2]. Our findings indicate that a fraction of distant icy minor planets can exhibit atmospheres possibly caused by ongoing cryovolcanic activity or a recent impact event of a small icy object.**

(612533) 2002 $XV_{93}$ (hereafter 2002 $XV_{93}$) is a plutino (a Kuiper belt object in 2:3 resonance with Neptune) with an orbit slightly larger than Pluto's in semi-major axis ($a$ = 39.4 au) and lower in eccentricity ($e$ = 0.13) and inclination ($i$ = 13.3°). Its absolute magnitude ($H \approx 5.4$) and thermal infrared observations with the Spitzer Space Telescope and Herschel Space Observatory indicate a mean radius of ~ 275 km and a very low albedo of ~ 4%[12]. On 10 January 2024 (UT), 2002 $XV_{93}$ was predicted to occult a Gaia G-band magnitude 15.8 star[13], as seen from regions spanning East Asia to Southeast Asia, including Japan (Methods).

Continued advances in occultation observation techniques have steadily enhanced the sensitivity to ever thinner atmospheres around TNOs[14,15]. High-sensitivity and high-time-resolution optical photometry enabled by recent advances in complementary metal-oxide semiconductor (CMOS) imagers can detect subtle refractive dimming features caused by tenuous atmospheric layers, even when using small telescopes accessible to citizen astronomers[16]. We carried out observations of the event at four stations across the predicted shadow path, partly as a framework for the observation campaign *Trans-Neptunian Atmospheres and Belts Analysis through Stellar-occultation Coordinated Observations* (TABASCO). Of the four planned stations, three obtained usable data under fine weather conditions (details in Table 1 and Methods); (1) Kyoto: a portable 20 cm telescope system with a commercial CMOS camera (SoCoSoCo PONCOTS[17]) operated from Kyoto University's rooftop; (2) Kiso Observatory: the 1.05 m f/3.1 Schmidt telescope equipped with the Tomo-e Gozen CMOS camera module[18]; and (3) Fukushima: a citizen astronomer-operated 25 cm telescope attached to a commercial CMOS camera. The Kyoto and Kiso sites' observations provide two positive occultation segments ("chords", Extended Data Fig. 1), whose four extremities are approximated by a circular shadow with the best-fit radius $R = 235^{+22}_{-15}$ km (Fig. 1 and Methods).

At the Kiso and Fukushima sites, we obtained detailed light curves probing the region immediately outside 2002 $XV_{93}$'s shadow edge, which are significantly different from those obtained in typical stellar occultations by atmosphere-free objects (Fig. 2). The light curve obtained at Kiso exhibited a similar gradual dimming of the flux at the edge of the shadow at both the immersion and emersion times of the occultation. The observed profiles clearly show a gradual flux decrease at the edge of the shadow, lasting for approximately 1.5 s. This behavior cannot be explained by the diffraction effect, given the Fresnel scale at the geocentric distance (~ 37 au) is ~ 1.2 km, corresponding to a duration of ~ 0.05 s, at a wavelength of 550 nm, or by the finite angular size of the occulted star (a diameter of ~ 0.0032 mas, corresponding to a duration of ~ 0.004 s; see Figs. 2a and b). On the other hand, the light curve at Fukushima shows no sharp drop of the stellar flux. Instead, a possible gradual drop with a duration of ~ 10 s was identified at ~ 4.0σ significance at the time of the closest approach to the shadow edge (Fig. 2c). Kiso's light curve as a function of the distance from the occultation center (Fig. 3) shows an abrupt change of slope, suggesting a thermal inversion caused by upper atmospheric heating of methane ($CH_4$), which is known for Pluto's atmosphere[19,20]. To fit to the Kiso data, synthetic light curves are thus generated using a ray-tracing technique by assuming a "Pluto-like" thermal profile (Methods). This process requires an assumption about the main atmospheric components. We explored two simplified models of atmospheric composition to enable straightforward comparison with previous investigations, including pure $CH_4$ and a Pluto-like atmosphere dominated by nitrogen ($N_2$). The latter case assumes a tiny fraction of $CH_4$, which can cause the upper atmospheric heating, but is negligible for the following model-fit approach. We should note that we cannot exclude the possibility that other volatile species could dominate the atmosphere since the present occultation light curves lack the capability to distinguish among different molecular species. For each composition hypothesis, we computed the refractivity profile using the adopted thermal structure and ray-traced the corresponding synthetic light curves (Methods). In these fits, the radius of the body $R$, surface pressure $p$, and the central occultation time are the free parameters, obtained by $\chi^2$ minimization to Kiso's light curve (Methods). The two best-fit synthetic light curves reproduce the observed profile (Fig. 3) with the best-fit $R$ of 244 ± 4 km. The best-fit surface pressure is $p = 124^{+15}_{-12}$ nbar and $p = 177^{+18}_{-19}$ nbar for the pure $CH_4$ and the $N_2$ cases, respectively (Extended Data Fig. 3). The $\chi^2$ values at the best fit are 28.7 and 28.5 with 29 degrees of freedom. Both represent an improvement compared to an atmosphere-free model that includes only diffraction and the stellar angular size, for which the corresponding fits yield $\chi^2 = 89.2$ with 30 degrees of freedom. The synthetic light curves with these pressures can also reproduce the possible gradual drop of the flux observed at the Fukushima station (Figs. 2c and 3), assuming a closest-approach distance of 230–240 km to the shadow center, which is consistent with that obtained from the chord map solution ($255^{+45}_{-43}$ km, see Fig. 1 and Methods). Due to its relatively low signal-to-noise ratio and the limited temporal sampling, the light curve at Kyoto cannot provide an independent constraint on the presence or absence of a gradual flux attenuation (Extended Data Fig. 2a). Nevertheless, the Kyoto data remains consistent with the synthetic light curves (Fig. 3). These best-fit pressures are approximately 50 to 100 times lower than the current surface pressure of Pluto observed between 1988 and 2020[6,7,16], and yet equivalent to or up to ~ 100 times above the previous upper limits obtained for other TNOs[8–11,15,21–23] (Fig. 4).

TNOs are remnants of the outer Solar System, composed largely of frozen volatiles.

However, despite targeted searches using stellar occultations, large TNOs other than Pluto have shown no evidence of atmospheres to date[3]. The dwarf planet (136199) Eris (radius ~ 1,163 km) was observed in a multi-chord occultation, which revealed no evidence of a global atmosphere. The observations place an upper limit of approximately 10 nbar on the surface pressure of any potential atmosphere composed of $N_2$, $CH_4$, or argon[8]. This indicates that (136199) Eris's volatiles are currently frozen out, which is consistent with its great distance at present (97 au, Fig. 4b). The other two dwarf planets in the Kuiper belt, (136108) Haumea (radius ~ 800 km) and (136472) Makemake (radius ~ 710 km), were expected to possibly harbor a Pluto-like atmosphere at their intermediate heliocentric distance (Fig. 4b) but demonstrated an absence of any global atmosphere down to the 50–100 nbar level[9,10]. This suggests that their modest supply of $CH_4$ and other volatiles is mostly sequestered on their surfaces in the form of ice. Similarly, the mid-sized TNO (50000) Quaoar (radius ~ 555 km) exhibits surface $CH_4$ ice but no measurable atmosphere with its surface pressure upper limit of ~ 1 nbar[11]. Additionally, recent stellar occultation observations of other 200–300 km sized TNOs, such as (28978) Ixion, (38268) Huya, and (84522) 2002 $TC_{302}$ have likewise yielded non-detections of atmospheres[21,15,22] with their heliocentric distances comparable to that of 2002 $XV_{93}$ (38 au, Fig. 4b).

These previous observational constraints have reinforced the understanding that TNOs smaller than approximately 1000 km cannot retain atmospheres over Gyr timescales. This consensus arises from theoretical considerations that volatile loss rates on such bodies are substantial, with the hydrodynamic or rapid thermal escape[1,2]. The expected ratio between the gravitational potential of a molecule near the surface and its thermal energy, which is referred to as the Jeans parameter $\lambda$, is

$$\lambda = \frac{GM\mu m_p}{Rk_B T} = \frac{4\pi G\rho R^2 \mu m_p}{3k_B T}, \quad (1)$$

where $G$ is the gravitational constant, $M$, $R$, and $\rho$ are the mass, radius, and bulk mass density of the TNO, and $\mu$ and $m_p$ are the mean molecular weight of the atmospheric gas and the mass of the proton, $k_B$ is the Boltzmann constant, and $T$ is the surface TNO temperature. For a $R$ ~ 250 km radius TNO with a Charon-like bulk mass density $\rho = 1.5 \times 10^3$ kg m$^{-3}$ and $T = 40 - 50$ K, $\lambda$ is close to unity. When $\lambda \sim 1$, the gases in the atmosphere hydrodynamically escape to interplanetary space at a timescale significantly shorter than the age of the Solar System[1,2,24]. Recent near-infrared spectroscopy with the James Webb Space Telescope has indeed revealed no prominent absorption features attributable to volatile-ice species on 2002 $XV_{93}$[25], consistent with a scenario in which atmospheric escape has depleted the majority of the surface's volatiles.

Sublimation of surface volatiles alone is therefore not expected to sustain a permanent atmosphere.
Possible alternative sources for the unexpected atmosphere include (1) outgassing associated with cryovolcanic activity, and (2) a recent impacting event of a small KBO or Oort Cloud object. Cryovolcanism is generally associated with the presence of internal melting materials. In fact, observational evidence for recent or ongoing geochemical evolution of larger TNOs such as (90377) Sedna, (225088) Gonggong, and (50000) Quaoar have been discovered and interpreted as the possible presence of a subsurface melting layer beneath their icy shells[26]. In general, for objects with radii of order ~ 250 km, the likelihood of long-lived and large-scale cryovolcanic activity is generally considered low, since their small sizes and limited internal heat budgets favor the development of thick, cold lithospheres that conductively cool over time[27]. Nevertheless, under special circumstances, such as unusually high abundances of antifreeze agents (e.g., ammonia or methanol) or tidal forcing from a satellite whose presence or absence could not be strongly constrained by the current occultation data, cryovolcanic-like seepage or venting could still occur[28].
A recent impact event is another plausible source of the observed atmosphere. Although such impact-generated gases would be ephemeral, as they quickly recondense or escape, they are thought to form a short-lived (on a timescale of less than $10^2$ years[29]) atmosphere around a several-hundred-km-sized TNO. Since the relative velocity between the impactor and a plutino is expected to have been very low[30], a significant fraction of the volatiles originating from its interior would likely have been retained around the target body rather than escaping into interplanetary space immediately after the impact. A comet-like impactor of radius a few $10^2$ m carrying relatively rich volatiles contains $10^{11}$ – $10^{12}$ kg of $CO/CH_4/N_2$[29] comparable to or greater than the amount required to account for a global ~ $10^2$ nbar scale atmosphere around 2002 $XV_{93}$. Also, volatiles (or a possible subsurface ocean) beneath a surface less-volatile layer could be excavated by such impacts and form a potential atmosphere[27].
2002 $XV_{93}$'s atmosphere is the first example of a sub-$\mu$bar scale atmosphere around a body other than large and dwarf planets and their large satellites in the Solar System. This discovery means that the traditional idea that global dense atmospheres only form around larger planets must be revised. These results also highlight the significant potential of collaboration between citizen and professional astronomers in conducting stellar occultation observations. Such observations complement those made using major facilities such as the James Webb Space Telescope, the Vera C. Rubin Observatory, and upcoming 10–30 m class telescopes. If the atmosphere is impact-generated, we predict a monotonic decline in surface pressure $p$ over the next several years[28]. Persistent or seasonal variations, however, would favor endogenous outgassing. Coordinated professional–citizen networks are uniquely positioned to obtain this definitive time series.

## Methods

### Prediction of the occultation

The stellar occultation by 2002 $XV_{93}$ on 10 January 2024 was predicted using an ephemeris for the object (JPL18 orbital elements) and a target star from the Gaia DR3 catalogue[13] (source ID: 940732910252854272, Gaia Gmag = 15.8). According to the Gaia DR3 catalogue, Re-normalised Unit Weight Error (RUWE) is 0.98, close to the normal value (RUWE ~ 1), and the duplicity flag is false, indicating that the occulted star is not a double. According to the prediction, 2002 $XV_{93}$'s shadow was expected to pass over East Asia, including Japan, at a relative velocity of $v_{rel}$ = 25.2 km $s^{-1}$ (Extended Data Fig. 1). This occultation event was also predicted by the European Research Council (ERC) *Lucky Star*[31].

### Observations

Table 1 summarizes the circumstances of each observing site. Out of four sites, including the three stations organized under the TABASCO campaign, three (Kyoto, Kiso, and Fukushima) succeeded in making photometric observations of the occulted star (See Extended Data Fig. 2).

At Kyoto, we observed the occultation with *SOlving COmet and other SOlar-system Complexity Planetary ObservatioN Camera for Optical Transient Surveys* (SoCoSoCo PONCOTS), which is a compact optical observation system based on the PONCOTS and OASES systems[17,32,33]. SoCoSoCo PONCOTS consists of the SONY IMX462 monochromatic CMOS sensor mounted on a 0.20 m f/2.0 prime focus astrograph. The exposure time for each image was set to be 1.0 s. The internal clock of the control PC was synchronized with the Global Positioning System (GPS) time information received from a GPS receiver.

At Kiso, photometric data were obtained using the Tomo-e Gozen camera mounted on a 1.05 m f/3.1 Schmidt camera[18]. Tomo-e is an optical wide-field and high time resolution camera equipped with 84 CMOS sensors. In the present observation, we used a single sensor pointed at the occulted star. The field of view (FoV) for the single sensor corresponds to 39'.7 × 22'.4 with an angular pixel scale of 1".19. The exposure time was set to be 0.5 s. The timestamps for the obtained images are synchronized with the GPS time information. We should note that we previously carried out the observation of the stellar occultation (Gaia G mag = 15.7) by a TNO (50000) Quaoar on 28 June 2019 (UT) with Tomo-e under the same gain and exposure conditions[14]. The obtained light curve of the occultation showed the instantaneous ingress and egress of the stellar flux with a timescale significantly faster than the exposure time (0.5 s), indicating that the Tomo-e detector has an effectively instantaneous response.

At Fukushima, a citizen astronomer K. Hosoi used the ZWO ASI290MM monochromatic CMOS camera mounted on a 0.25 m f/4.0 Newtonian telescope. The exposure time for each image was set to be 0.49 s. For the timing calibration, the one pulse per second (1 PPS) emission of the Light-Emitting Diode (LED) generated by the GPS module and projected onto the image before and after the expected occultation time was used for the timing calibration. The timing was corrected by subtracting the offset between the original timestamp and the 1 PPS signal recorded in the image[34].

## Data reduction

Standard aperture photometry is performed for the images of the occulted star. Due to possible rapid changes in atmospheric conditions, the obtained flux values can suffer time-dependent fluctuations. To correct the possible flux fluctuations, we carry out differential photometry with nearby stars simultaneously detected in the same sensor. The calibrated light curve is then normalized to the unocculted flux. Extended Data Fig. 2 shows the normalized light curves obtained at the three observation sites.

From the three observation sites, this campaign resulted in two positive detections at the Kyoto SoCoSoCo PONCOTS and the Kiso Tomo-e Gozen. The apparent immersion and emersion times of the occultation were measured by the fit of the Fresnel diffraction fringe of an infinite, straight edge of a shadow convolved by the angular diameter and the spectral energy distribution of the star, and the efficiency and the finite integration time of the sensor, to the observed light curve[14]. The apparent angular diameter of the star was assumed to be 0.0032 milliarcseconds, based on the parameters provided in the Gaia DR3 catalogue[13]. The obtained chord durations of the occultation at Kyoto and Kiso are 18.0±0.9 and 16.53±0.08 s, which correspond to positive chord lengths of 458±22 and 422±2 km, respectively. Since we have only two positive chords that have similar median lines in the chord plot, we fit a circular shadow to the $N = 4$ chord extremities by finding the minimum $\chi^2$ defined as

$$\chi^2 = \sum_i^N \frac{(r_{i,obs} - R)^2}{\sigma_{r,i}^2}, \qquad (2)$$

where $(r_{i,obs} - R)$ is the difference between the distance of the observed chord extremity $i$ from the model shadow center $(f_c, g_c)$ and the model circular shadow radius $R$, and $\sigma_{r,i}$ is the $1\sigma$ uncertainty in the radial direction. This fit thus has 3 free parameters ($f_c$, $g_c$, and $R$). Fig. 1 is a chord map of the occultation event in the projected sky plane based on the results of the three-station observations overlaid with the best-fit circular shadow. The observed chords are approximated by a circular shadow with a best-fit radius of $235^{+22}_{-15}$ km, which is slightly smaller than the estimated radius obtained by the previous space-borne mid- to far-infrared observations ($275^{+11}_{-12}$ km[12]). The best-fit $\chi^2$ value is $\chi^2 = 1.40$ for $N - M = 1$ degree of freedom, and the assumption of a circular shadow appears to be valid in the present case. The expected closest approaches of the Kyoto, Kiso, and Fukushima stations to the shadow center are measured to be $41^{+45}_{-41}$, $103^{+45}_{-43}$ km and $255^{+45}_{-43}$ km, respectively. The shadow center inferred from the fit differs from the predicted path according to both the pre-event forecast (Extended Data Fig. 1) and an updated ephemeris. The best fit $(f_c,\ g_c)$ relative to its predicted position with the latest JPL24 ephemeris as of September 2025 is $(f_c,\ g_c) = (-315^{+3}_{-4}\ \mathrm{km}, 277^{+43}_{-44}\ \mathrm{km})$.

## Atmosphere refraction model fit

Since dynamical, chemical, and thermophysical properties of 2002 $XV_{93}$ are highly uncertain, it is difficult to construct a detailed atmosphere refraction model. For a radial profile of 2002 $XV_{93}$'s lower atmosphere, we assume a spherically symmetric, Pluto-like hydrostatic atmosphere with a temperature inversion layer, following the previous TNOs' atmosphere studies[11,14,15,21]. In this model, the temperature profile increases from the equilibrium surface temperature up to an altitude $z$ of $z \sim 20$ km with a gradient of $dT/dz \sim 2.7$ K km$^{-1}$. For the surface temperature at $z = 0$, we adopt $T = 47$ K, which corresponds to the average surface temperature derived from the model thermal parameters obtained from the previous infrared observations[12] and the Near-Earth Asteroid Thermal Model (NEATM)[35]. The number density, $n(z)$, is derived by integrating the hydrostatic equation for ideal gas upwards from the surface,

$$\frac{dp(z)}{dz} = -\mu\, n(z)\, g(z), \tag{3}$$

where $p(z)$ is the pressure at $z$, $\mu$ is the mean molecular mass, and $g(z)$ is the gravitational acceleration under the assumption that the bulk mass density of 2002 $XV_{93}$ to be $1.5 \times 10^3$ kg m$^{-3}$. The Jeans parameter at surface of 2002 $XV_{93}$ is expected to be $\lambda \sim 1$ (see text). With such lower $\lambda$ conditions, the density profile at higher altitudes ($z \gtrsim 25$ km) is thus approximated with the free molecular flow profile[24], i.e., $n(z) \propto (z + R)^{-2}$. We note that the upper atmospheric density profile does not significantly affect the resultant synthetic light curves provided that it is smoothly connected to the lower profile.

The atmospheric refractivity $\nu(z)$ at $z$ is derived by $\nu(z) = n(z)\,K$, where $K$ is the molecular refractivity. The bending angle $\omega(r)$ of a stellar ray with an impact parameter $r$ from the shadow center is

$$\omega(r) = \int_{-\infty}^{+\infty} \frac{r}{x}\frac{\partial \nu(z)}{\partial r}\,dx, \tag{4}$$

where $x$ is the ray trajectory. The distance of the ray from the shadow center in the observer plane $y$ is $y = r + \omega(r)\,D$, where $D$ is the distance of the shadow ($D = 37.0$ au). To convert the ray-bending field into a model light curve, we map each impact parameter $r$ to its location in the observer plane $y$. Since we assume a transparent, spherically symmetric atmosphere, the normalized synthetic stellar flux $\phi_{\mathrm{syn}}(y)$ is then

$$\phi_{\mathrm{syn}}(y) = \frac{dr}{dy}\frac{r}{y} = \frac{1}{1+Dd\omega/dr}\,\frac{1}{1+D\omega/r}, \tag{5}$$

where the second term corresponds to the focusing factor caused by the circular limb curvature[36].

The synthetic stellar flux is fitted to the data points of the immersion/emersion times of the occultation observed at Kiso by minimizing the following $\chi^2$,

$$\chi^2 = \sum_i^N \frac{(\phi_{i,\mathrm{obs}} - \phi_{i,\mathrm{syn}})^2}{\sigma_i^2}, \tag{6}$$

where $\phi_{i,\mathrm{obs}}$ and $\phi_{i,\mathrm{syn}}$ are the observed and synthetic stellar fluxes at a data point $i$ ($N$ data points in total), and $\sigma_i^2$ is the $1\sigma$ error of $\phi_{i,\mathrm{obs}}$. The best-fit results are shown in Figs. 2a, 2b, and 3. The $\chi^2$ maps as for the two composition cases ($CH_4$ and $N_2$ dominant) are shown in the Extended Data Fig. 3.

To assess whether the best-fit atmospheric profile derived from the Kiso data is consistent with the Fukushima observation, we also attempted a fit to the Fukushima light curve. In principle, fitting an atmospheric model to a chord without detection of the body shadow requires precise knowledge of the distance from the shadow center and of the shadow oblateness. However, with only four extremities available from the present campaign, these quantities cannot be tightly constrained. Therefore, instead of performing an independent fit, we adopted the best-fit atmospheric parameters obtained from the Kiso data and generated the corresponding synthetic refractive light curve. In this procedure, the closest approach distance of the Fukushima chord to the shadow center was treated as the sole free parameter, and $\chi^2$ minimization was carried out accordingly. The best-fit results are shown in Figs. 2c and 3, and Extended Data Fig. 2c.

**Data availability**

The observational data used in this study are available at Zenodo.

**Code availability**

Codes written for this study are available from the corresponding author upon request.

**Acknowledgements**

This research has been partly supported by Japan Society for the Promotion of Science (JSPS) Grants-in-Aid for Scientific Research (KAKENHI) grant Nos. 18K13606 and 21H0115.This research is partially supported by the Optical and Infrared Synergetic Telescopes for Education and Research (OISTER) program funded by the MEXT of Japan. The TABASCO campaign is partially supported by the International Occultation Timing Association – East Asia (IOTA/EA).

**Author contributions**

K.A. planned and organized the campaign, made the occultation prediction, participated in the observations, obtained and analyzed the data, and interpreted the data and wrote the paper. F.Y. helped organize the campaign, interpret the data, and write the paper. B.S. helped to plan the campaign. S.T. participated in the observations, obtained and analyzed the data, and helped write the paper. K.H. participated in the observations, obtained and analyzed the data. T.O. and J.W. helped interpret the data and write the paper.

**Table 1: Circumstances of the 10 January 2024 stellar occultation observations for observing stations**

| Observation site | Longitude<br>Latitude<br>Altitude | Aperture<br>Camera | Status | Observer |
|---|---|---|---|---|
| Kyoto<br>(Kyoto University) | 135° 46' 59".9 E<br>35° 01' 51".2 N<br>62 m | 0.20 m<br>SoCoSoCo<br>PONCOTS | Positive | K. A. |
| Kiso<br>(Kiso Observatory) | 137° 37' 31".5 E<br>35° 47' 50".0 N<br>1,132 m | 1.05 m<br>Tomo-e Gozen | Positive | S. T. |
| Fukushima | 140° 26' 04".2 E<br>37° 25' 36".7 N<br>274 m | 0.25 m<br>ASI290MM | Negative | K. H. |
| Okayama<br>(Okayama Observatory) | 133° 35' 48".2 E<br>34° 34' 36".8 N<br>355 m | 3.78 m<br>TriCCS | Bad weather<br>(No obs.) | K. A. |

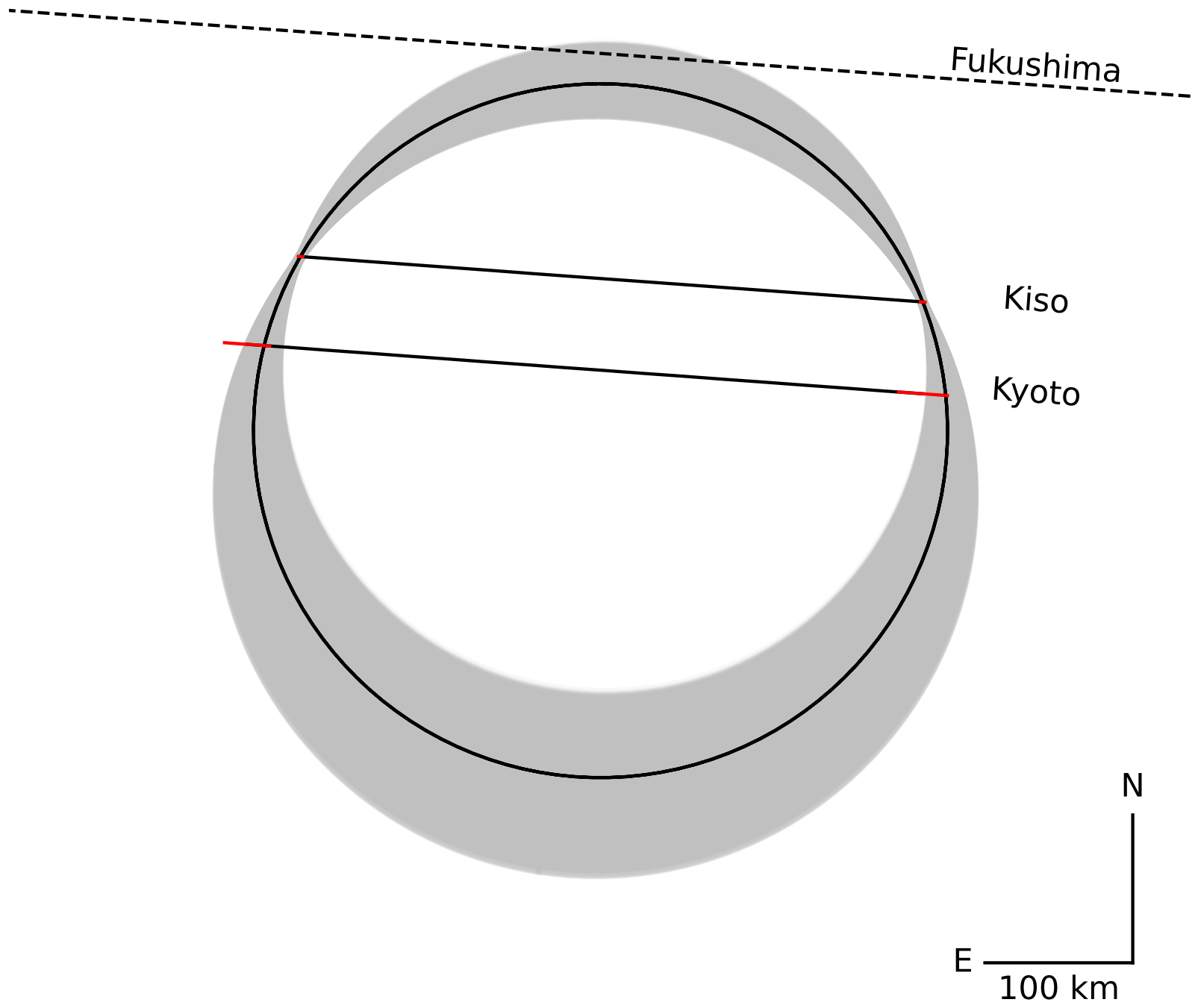

**Figure 1: Sky-plane projection of the 2002 XV$_{93}$ stellar occultation detections.** The solid black circle is the best-fit solution of the 2002 XV$_{93}$'s circular shadow obtained with the four extremities of the two chords (Kyoto and Kiso, solid lines). The red lines in the extremities of the chords correspond to uncertainties of the immersion and emersion times. The gray region represents 1$\sigma$ uncertainty of the circular shadow. The dashed line is the negative chord at Fukushima. The celestial north and east directions with respect to the J2000 equinox are shown in the lower right corner.

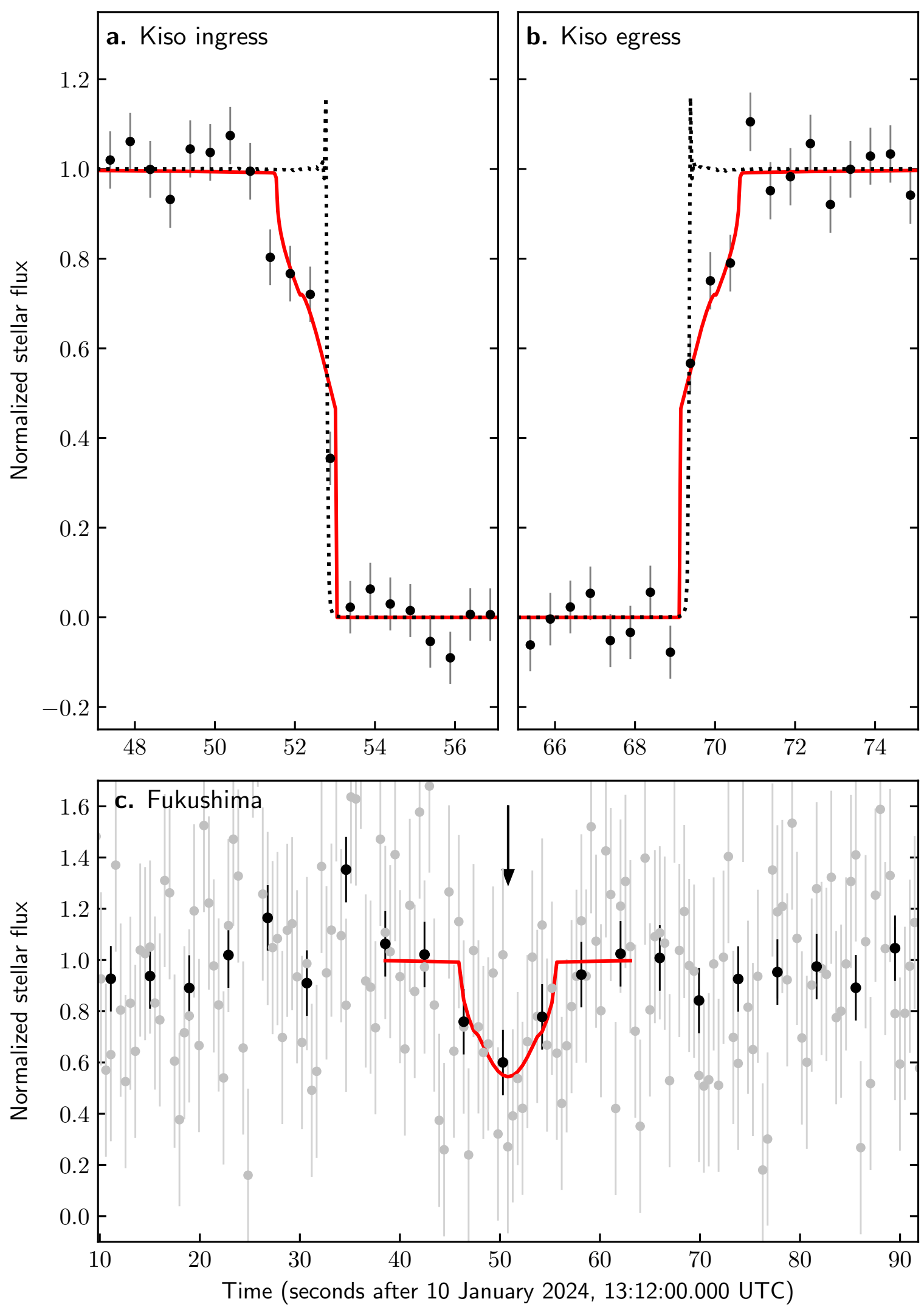

**Figure 2: Light curves of the possible atmospheric features.** The observed stellar flux points of the 2002 $XV_{93}$ stellar occultation (points with error bars) on 10 January 2024 are overlaid with the example of the synthetic light curves (red lines), which assume a pure $CH_4$ atmosphere and a best-fit surface pressure of 124 nanobars. **a** and **b**, The ingress and egress light curves observed by the Tomo-e Gozen camera on the Kiso 1.05 m f/3.1 Schmidt telescope (Kiso station). The dotted lines represent the synthetic light curves with no atmosphere (see Methods). **c**, The light curve obtained by a citizen astronomer K. Hosoi (Fukushima station). The gray points are photometry for single exposures, and the black points correspond to the flux points binned by a factor of 8. The vertical arrow represents the closest approach time to the shadow edge at Fukushima, at 13:12:50.818 UT.

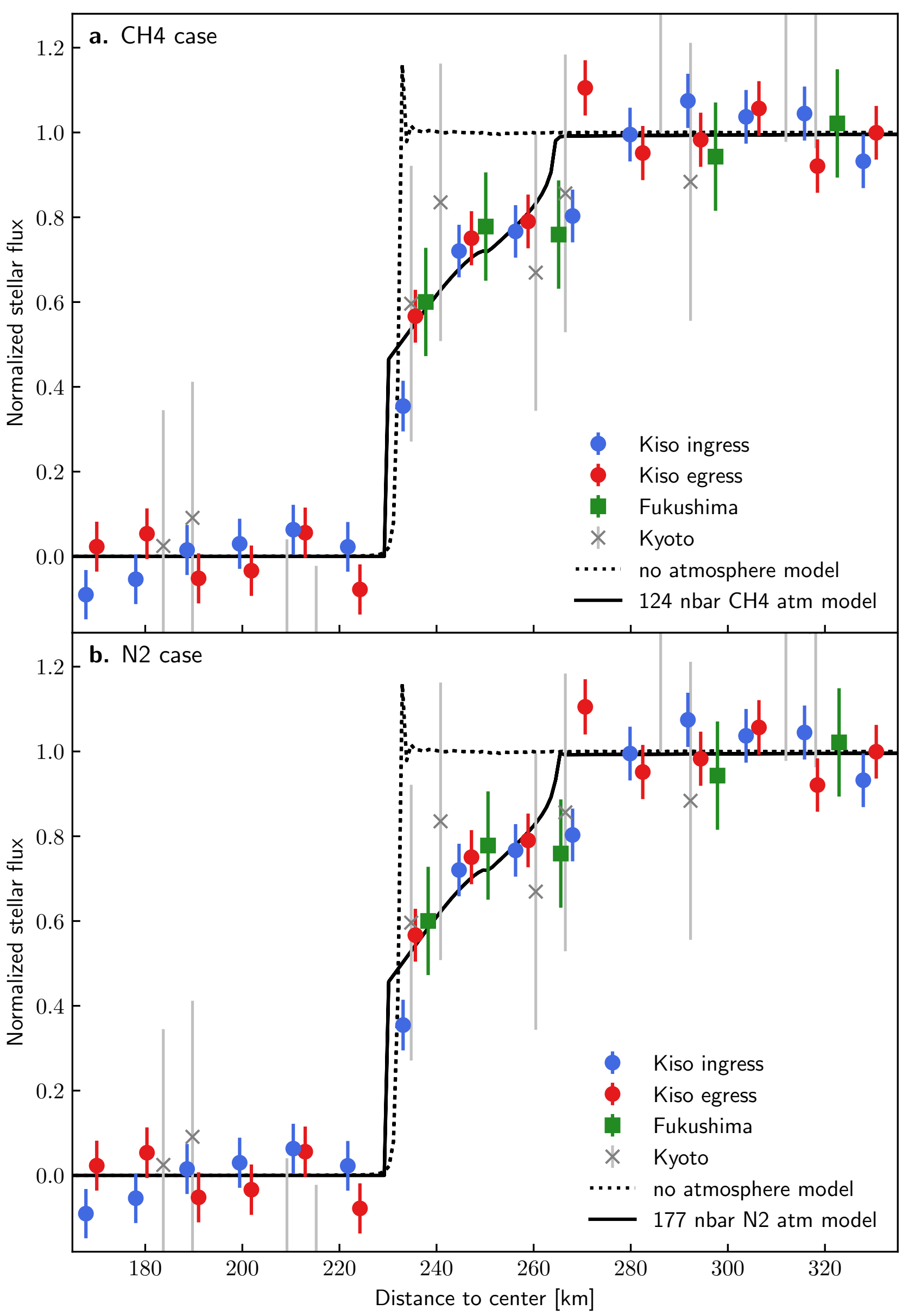

**Figure 3: 2002 $XV_{93}$'s atmospheric refraction profile.** The flux data points (Kiso ingress; blue circles, Kiso egress; red circles, Fukushima binned photometry; green squares; Kyoto; gray crosses) are plotted as a function of the distance from the 2002 $XV_{93}$'s center obtained with the circular shadow solution shown in Fig. 1. The solid lines are the best-fit model (**a**; pure $CH_4$, **b**; the Pluto-like $N_2$ models), and the dotted lines represent the synthetic profile with an atmosphere-free case.

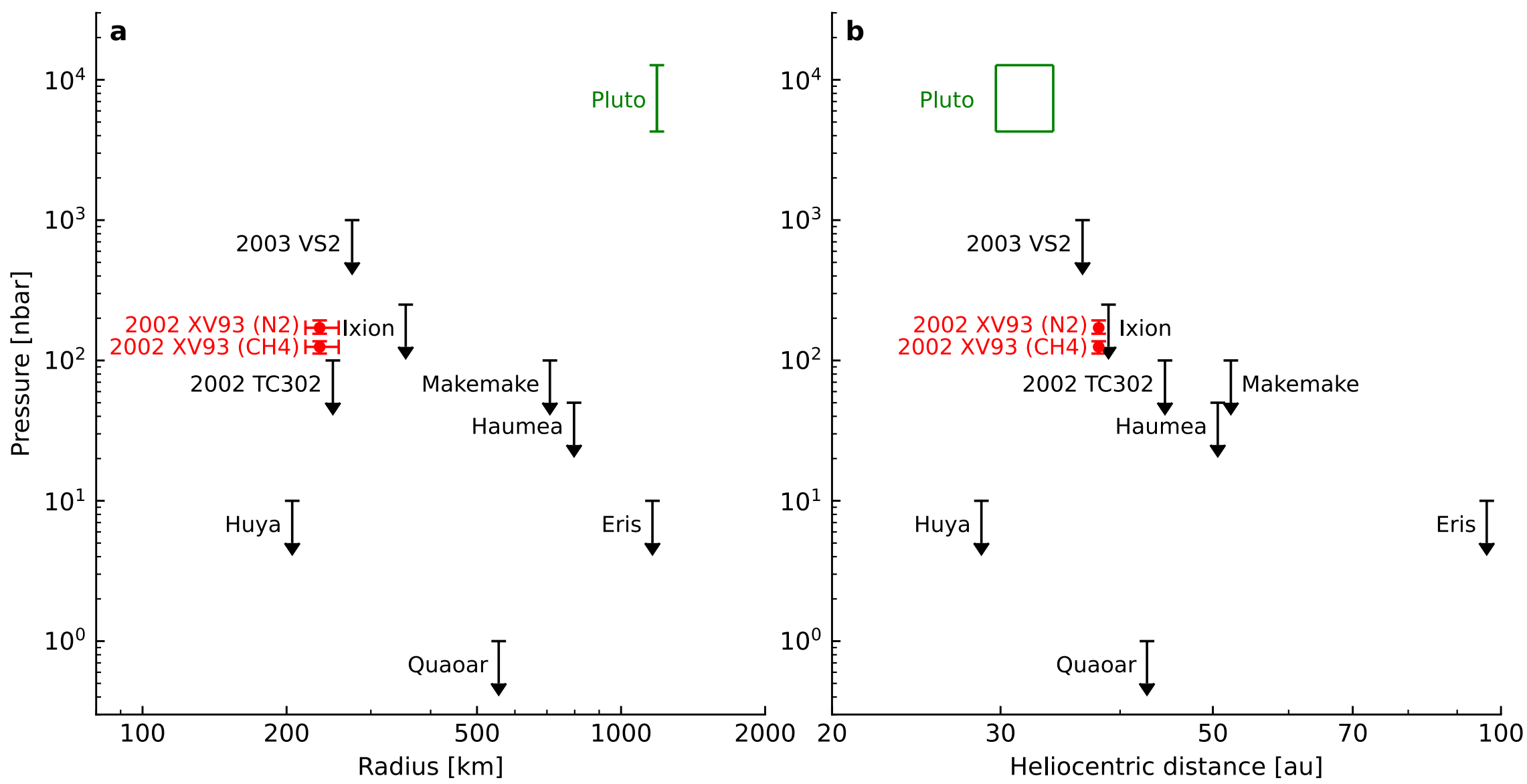

**Figure 4: Atmospheric pressure constraints on TNOs.** The constraints on the surface atmospheric pressure derived from stellar occultation observations are shown as a function of the radius of the TNOs (**a**) and of the heliocentric distance at the time of observations (**b**). The red points with error bars are the best-fit surface pressure constraints for 2002 $XV_{93}$ in cases with a $CH_4$ and $N_2$ atmosphere. The green line and box show the range of surface pressures on Pluto between 1988 and 2020[6,7,16]. The black arrows denote 3$\sigma$ upper limits derived from previous stellar occultation studies for the $CH_4$ atmosphere case[8–11,15,21–23].

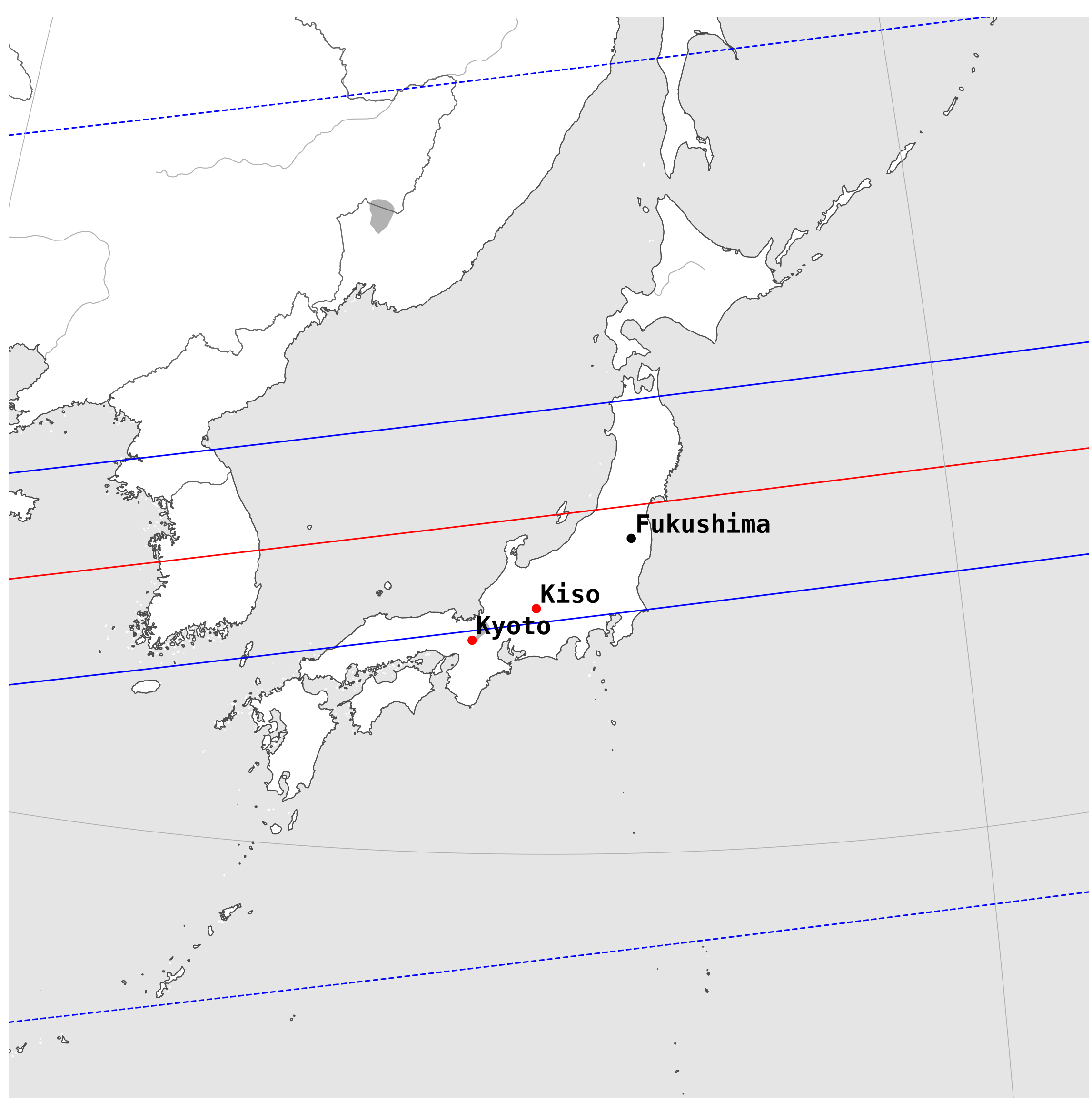

**Extended Data Fig. 1: Prediction map of the 2002 XV$_{93}$ stellar occultation event on 10 January 2024.** The red line shows the predicted central path of the occultation shadow, calculated using the JPL18 orbital elements and the Gaia DR3 stellar catalogue, and the blue solid lines correspond to the shadow width assuming a diameter for 2002 XV$_{93}$ of 275 km. The blue dashed lines represent the 1$\sigma$ error range. The red circle indicates the location of the positive detection (Kyoto and Kiso). The black point represents the Fukushima site, where the observed light curve shows no positive detection of occultation by the main body.

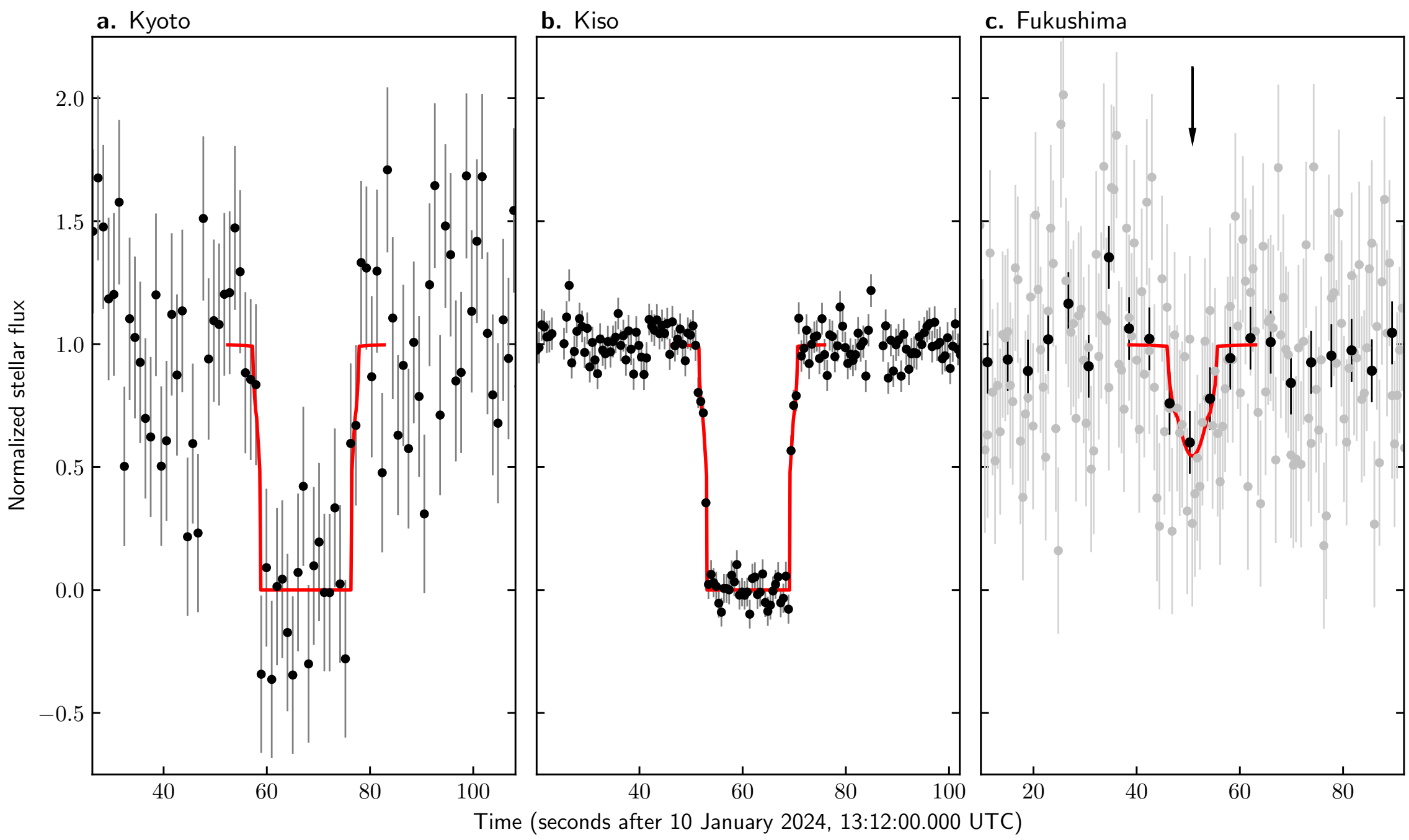

**Extended Data Fig. 2: Three station light curves of 2002 $XV_{93}$'s stellar occultation on 10 January 2024.** The flux points (points with error bars) observed at Kyoto (**a**), Kiso (**b**) and Fukushima stations (**c**) overlaid with the synthetic light curves (red lines, assuming a pure $CH_4$ atmosphere and a best-fit surface pressure of 124 nanobars) are shown as a function of time. In panel c, the gray points are the flux points for the individual exposures, and the black points are those binned by a factor of 8. The vertical arrow represents the closest approach time to the shadow edge at Fukushima, at 13:12:50.818 UT.

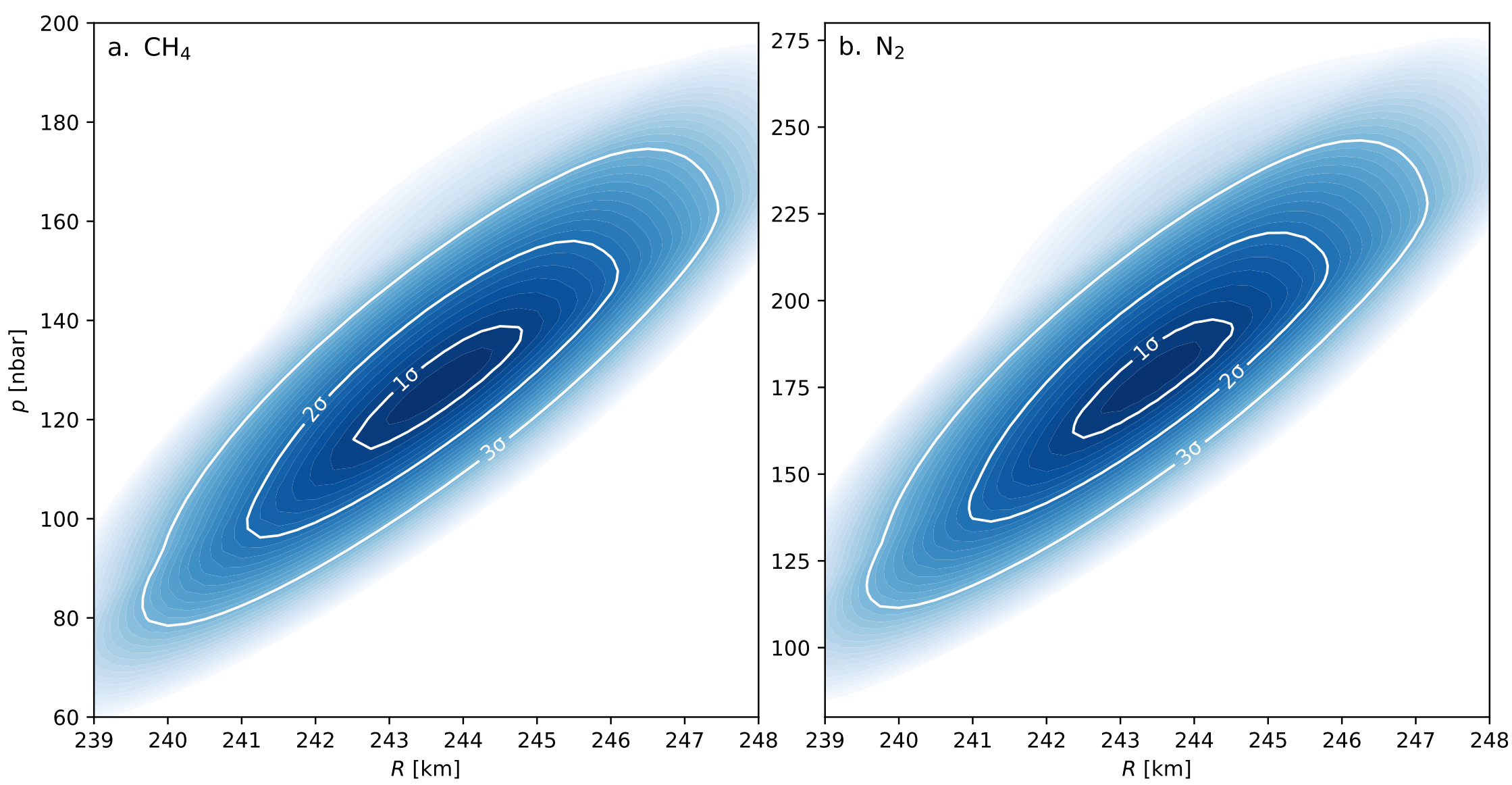

**Extended Data Fig. 3: $\chi^2$ map for the light curve fit**. The distribution of $\chi^2$ values as a function of the circular shadow radius $R$ and surface pressure $p$ obtained by the light curve fit to the pure $CH_4$ (**a**) and $N_2$ dominant (**b**) atmosphere refraction model.